\documentclass[cameraready]{Interspeech}

\title{Enhancing BEST-RQ Pseudo-Label Quality through Online Refinement for Automatic Speech Recognition}

\author[affiliation={1,2}, orcid=0009-0005-2872-5470]{Jingjing}{Xu}
\author[affiliation={1}, orcid=0009-0008-5302-3579]{Zijian}{Yang}
\author[affiliation={1,2}, orcid=0009-0003-3309-2538]{Mohammad}{Zeineldeen}
\author[affiliation={1,2}, orcid=0000-0003-1641-6628]{Eugen}{Beck}
\author[affiliation={1,2}]{Ralf}{Schlüter}
\author[affiliation={1,2}]{Hermann}{Ney}

\address{
    $^1$ Machine Learning and Human Language Technology Group, RWTH Aachen University, Germany \\
    $^2$ Apptek GmbH, Aachen, Germany
}

\email{\{jxu, zyang, zeineldeen, schlueter\}@ml.rwth-aachen.de, ebeck@apptek.com}


\usepackage{comment}
\usepackage{amssymb}
\usepackage{amsmath,graphicx,hyperref}
\usepackage{multirow}
\usepackage{array}
\usepackage{graphicx}
\usepackage{makecell}
\usepackage{pifont}
\usepackage{xcolor}    
\usepackage[normalem]{ulem} 
\usepackage{enumitem} 
\usepackage{etoolbox}
\usepackage{caption}
\newcolumntype{C}[1]{>{\centering\arraybackslash}p{#1}}

\graphicspath{{figures/}}

\def\LBS{{\textit{Librispeech}}}

\begin{document}

\maketitle

\begin{abstract}
  BEST-RQ is a simple and effective self-supervised training method for speech representation
  learning that performs well on automatic speech recognition (ASR) tasks.
  It generates pseudo-labels using a fixed online quantization scheme, which simplifies training but provides weaker supervision than HuBERT-style models that
  iteratively refine pseudo-labels.
  In this work, we improve online pseudo-label generation while preserving simplicity.
  We propose three modifications: replacing the quantizer’s
  linear projection with Principal Component Analysis (PCA), updating the codebook
  via iterative codebook refinement and introducing an additional codebook updated via codebook distillation.
  We pre-train on the {\LBS} 960-hour dataset and fine-tune using 100 hours of
  supervised {\LBS} data.
  With all three modifications enabled, we achieve a 12\% relative reduction in word error rate (WER) on the {\LBS} test-other set, improving from 10.1\% to 8.8\%.
\end{abstract}

\section{Introduction}
Self-supervised learning (SSL) can learn rich and generalizable representations from large
amounts of unlabeled data by generating pseudo-labels from the data itself \cite{gui2024sslsurvey}.
In recent years, different SSL approaches have been proposed for the
speech domain and achieved state-of-the-art performance on the
downstream ASR tasks \cite{schneider2019wav2vec, baevski22wav2vec2,hsu2021hubert,chung2021w2vbert,baevski2022data2vec,baevski2023data2vec2,chiu2022bestrq, chen2022wavlm}.

Wav2Vec 2.0 \cite{baevski22wav2vec2} uses a contrastive learning objective, in
which the model learns contextualized audio embeddings by distinguishing the true
latent representations from distractor samples.
One of its key limitations is the sensitivity to the selection of
distractors.
HuBERT \cite{hsu2021hubert} uses a BERT-style masked prediction loss and generates
pseudo-labels via iterative clustering of MFCC features or intermediate layer
representations.
While HuBERT achieves performance on par with or better than Wav2Vec 2.0,
its multi-round clustering requires additional computational resources, and
storing the pseudo-labels demands increased disk space.
Data2Vec \cite{baevski2022data2vec,baevski2023data2vec2,liu2023dinosr} adopts
a teacher-student
self-distillation framework, in which the student predicts the teacher’s latent
representations from partially masked inputs.
However, a notable drawback of Data2Vec is its high computational cost, along
with the need for careful hyper-parameter tuning.

The recently proposed BEST-RQ \cite{chiu2022bestrq} efficiently generates high-quality speech
representations at lower computational cost.
It uses masked prediction loss like HuBERT but replaces its iterative clustering with a fixed random-projection quantizer for
online pseudo-labeling.
Despite this simplification, BEST-RQ remains competitive on the downstream ASR tasks
and therefore has gained growing attention \cite{zhang2023googleusm,dubey2024llama,das2024speechverse,yang2025mbestrq,gaur2024ASTRA}.

The main limitation of BEST-RQ is its static quantizer, which is
randomly initialized and kept fixed throughout training.
The quantizer generates pseudo-labels from log-Mel features, which are low-level
speech representations with limited discriminability, resulting in weak training targets.
Moreover, as observed in \cite{chiu2022bestrq}, pseudo-label quality is highly sensitive to the random initialization of the codebook and projection matrix, leading to variations in pretraining quality even under identical settings.
While \cite{zhang2023googleusm} attempts to address this issue by using multiple
codebooks, it increases model complexity and computational cost.
To address this weakness, we aim to refine the pseudo-label generation process of
BEST-RQ while preserving its core simplicity.
We replace the quantizer’s linear projection with PCA and
iteratively refine the codebook during training, which together better align
the codebook with the data distribution and reduce sensitivity to random
initialization.
Our results show that with these two modifications on a single codebook, we can
achieve the same performance as using six random codebooks while reducing training
time by 45\%.
In addition, we introduce an additional codebook, distilled from intermediate layer representations that capture richer linguistic cues, which further improves target discriminability.
Taken together, with all three proposed methods,  we can reduce WER by
roughly 12\% relative (from 10.1\% to 8.8\%) on {\LBS} test-other when fine-tuning
on {\LBS} 100h, yielding substantial gains at minimal additional training cost.

\section{Online Pseudo-Label Refinement}
\subsection{BEST-RQ}
BEST-RQ \cite{chiu2022bestrq} trains a model to predict masked speech segments using discrete
labels derived from a fixed random projection-based quantizer.
Specifically, the quantizer maps speech inputs through a randomly initialized
projection matrix and assigns them to the nearest entry in a randomly generated
codebook.
Both the projection matrix and the codebook remain fixed and are not updated
throughout self-supervised training.
This is then followed by BERT-style pre-training \cite{devlin2019BERT},
in which the masked speech inputs are processed by a Conformer-based encoder \cite{gulati2020conformer} and the model is optimized using a masked prediction loss.


\subsection{Our proposed approach}
\begin{figure}
  \centering
  \includegraphics[scale=0.38]{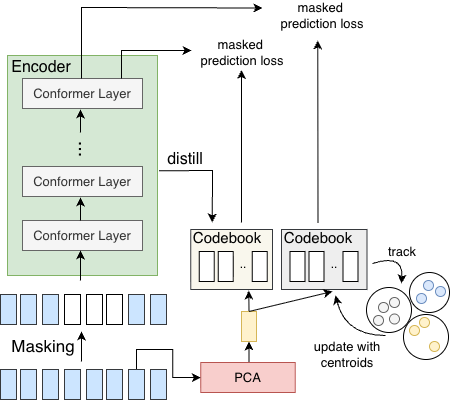}
  \caption{Proposed modifications to BEST-RQ: replacing the quantizer’s linear
  projection with PCA, updating the codebook via iterative codebook refinement, and adding an
  additional codebook updated via codebook distillation.}
  \label{fig:enhanced_best_rq}
  \vspace{-1.5em}
\end{figure}

In this section, we introduce three refinement methods for BEST-RQ, which improve
the quality of online targets from a fixed quantizer while preserving the
simplicity of online generation, as illustrated in Figure~\ref{fig:enhanced_best_rq}.

\subsubsection{PCA projection}
PCA projects features into a lower-dimensional
space while preserving as much variability as possible.
To retain the most informative structure of the input features, we replace the
random linear down-projection $W \in \mathbb{R}^{F \times F'}$, with $F'<F$ by PCA.
Furthermore, to avoid offline PCA computation and maintain a simple pipeline,
we adopt an incremental PCA method \cite{ross2008incremental} that can estimate the components online during
training.
For a new batch, we denote the speech input as $X_b \in \mathbb{R}^{N_b \times F}$,
where $N_b$ is the total number of frames in that batch and $F$ is the number of channels.
Let $\mu_{b-1} \in \mathbb{R}^{1 \times F}$, $\Sigma_{b-1} \in \mathbb{R}^{F \times F}$,
  $V_{b-1} \in \mathbb{R}^{F \times F}$ denote
the accumulated mean,  diagonal matrix of singular values and singular vectors from previous step.
$N_{\text{seen}}$ is the total number of frames seen so far in training.
We update the mean incrementally as:
\scalebox{0.9}{\parbox{1.1\linewidth}{%
\begin{align}
\bar{x}_b = \frac{1}{N_b} \sum_{i=1}^{N_b} X_b^i, \quad \mu_b = \frac{N_{\text{seen}} \mu_{b-1} + N_b \bar{x}_b}{N_{\text{seen}} + N_b}
\end{align}
}}
The correction term for the previous batches is:\\
\scalebox{0.9}{\parbox{1.1\linewidth}{%
\begin{align}
\Delta_b = \sqrt{\frac{N_{\text{seen}} N_b}{N_{\text{seen}}+N_b}} (\bar{x}_b - \mu_{b-1})
\end{align}
}}
The augmented matrix for singular value decomposition (SVD) is then:\\
\scalebox{0.9}{\parbox{1.1\linewidth}{%
\begin{align}
X'_b =
\begin{bmatrix}
\Sigma_{b-1} V_{b-1}^T \\
X_b - \bar{x}_b \\
\Delta_b
\end{bmatrix},
\quad U_b \Sigma_b V_b^T = \text{SVD}(X'_b)
\end{align}
}}
we replace the $W$ with top $F'$ right singular vectors from $V_b^T$ obtained via SVD of $X'_b$.
In practice, we stop updating $V_b^\top$ after one epoch.

\subsubsection{Iterative codebook refinement}
We also iteratively refine the random codebook during training by updating each codebook entry
with the corresponding centroid.
For each codebook entry $C_i, i=\{1,..,|C|\}$ in the codebook $C$, we have two
accumulators $s_i$ and $n_i$, both are initialized as 0 at the beginning
of each iteration.
\begin{itemize}[left=0.5pt, itemsep=1pt, topsep=2pt, parsep=0pt]
  \item $s_i$: the sum of the projected features assigned to $C_i$
  \item $n_i$: the count of assignments to $C_i$.
\end{itemize}
For each time frame t in the training data, the projected speech input $\hat{x}_t$ is assigned to its
nearest codebook entry $i^\ast = \arg\min_{i \in C} \lVert \hat{x}_t - C_i \rVert_2$,
we then update the two accumulators $s_{i^\ast} \leftarrow s_{i^\ast} + \hat{x}_t$,
$n_{i^\ast} \leftarrow n_{i^\ast} + 1$.
After a fixed number of updates, we update each codebook entry using the corresponding centroid:\\
\begin{equation}
C_i \leftarrow \frac{s_i}{n_i},
\quad \forall i \in \{1, \dots, |C|\}.
\end{equation}

\subsubsection{Codebook distillation}
In HuBERT \cite{hsu2021hubert}, pre-training uses an iterative pseudo-labeling
approach.
The first iteration applies k-means clustering to MFCC features, while the second
iteration clusters intermediate layer representations to produce refined labels,
improving downstream ASR performance.
This finding suggests that the intermediate representations capture richer
information than raw speech input features and could be leveraged to refine
pseudo-labels.
\cite{huo2025cmp} finds that iterative refinement of pseudo-labels is crucial in
guiding HuBERT to learn richer linguistic representations.
An open question is how to exploit this information in an online fashion for BEST-RQ.

Directly clustering intermediate representations is problematic because the model
parameters change rapidly during training, causing the representations to be
unstable as anchors.
However, while the representations themselves fluctuate, the temporal
relationships between frames remain relatively stable.
Motivated by this observation, we propose to distill the temporal structure of
the hidden representations into the codebook.
To explicitly capture temporal structure, we use temporal self-similarity
matrices, which measure the pairwise similarity between all frame
vectors across time.
Let $\hat{X} = [\hat{x}_1, ..., \hat{x}_T ]$ denote the speech input sequence
after linear projection.
For hard assignment, each frame $\hat{x}_t$ is represented by the closest codebook
entry $C_{i^*}$ in Euclidean distance, i.e. $\tilde{x}_t = C_{i^{*}}$.
For soft assignment, we first compute cosine similarity between $\hat{x}_t$ and codebook entry $C_i$.
\scalebox{0.9}{\parbox{1.1\linewidth}{%
  \begin{align}
  s_{t,i} = \frac{\hat{x}_t \cdot C_i}{\lVert \hat{x}_t\rVert \lVert C_i\rVert} \quad
  \alpha_{t,i} = \frac{\exp(s_{t,i})}{\sum_{j=1}^{|C|} \exp(s_{t,j})}
  \end{align}
}}
The frame representation $\tilde{x}_t$ is then obtained as a weighted sum
over all codebook entries.
$\tilde{x}_t=\sum_{i=1}^{|C|} \alpha_{t,i} \, C_i$.
The temporal self-similarity over reconstructed frame vectors is:
\scalebox{0.9}{\parbox{1.1\linewidth}{%
  \begin{align}
  S_C = \text{norm}_{\ell_2}(\tilde{X}) \, \text{norm}_{\ell_2}(\tilde{X})^\top,
  \end{align}
}}
where $\text{norm}_{\ell_2}$ normalizes each row to have unit $\ell_2$ norm.
Similarly, let $H_l \in \mathbb{R}^{T \times F}$ denote $l^{th}$ layer output representation.
Its temporal self-similarity matrix is defined as:
\scalebox{0.9}{\parbox{1.1\linewidth}{%
  \begin{align}
  S_H = \text{norm}_{\ell_2}(H_l) \, \text{norm}_{\ell_2}(H_l)^\top
  \end{align}
}}
Finally, to distill the temporal structure of the intermediate representations
into the codebook, we minimize the element-wise absolute difference between the
similarity matrices:
\scalebox{0.9}{\parbox{1.1\linewidth}{%
  \begin{align}
  \mathcal{L}_\text{distill} = \sum_{i=1}^{T} \sum_{j=1}^{T} \big| (S_H)_{ij} - (S_C)_{ij} \big|
  \end{align}
}}
This encourages the pseudo-labels to better approximate those from intermediate
layer outputs, providing a more informative training signal.
In practice, we apply distillation only to the unmasked frames, as they provide
a more reliable representation.
Furthermore, we enable distillation after approximately 30\% of the training
process, ensuring that the model has already reached a reasonable level of stability.

\section{Related Work}
The authors of \cite{zhang2023googleusm} propose a simple
approach that uses multiple codebooks instead of a single one.
Specifically, given N codebooks, N independent quantization targets are generated,
and the model employs N softmax layers to predict them.
The loss for each prediction is then scaled by a factor of $\frac{1}{N}$.
This formulation effectively ensembles the pseudo-labels, providing a richer,
more stable self-supervised signal, and it further reduces sensitivity to the
random initialization of a single codebook.
Building on this idea, the authors of \cite{baumann2025bestrq} improve the approach by using  multiple codebooks per cluster obtained from low-level feature clustering.
More recently, \cite{jiang2025birq} further extends BEST-RQ by discretizing intermediate representations to form enhanced pseudo-labels, while raw input derived anchoring labels provide a stable supervision signal and prevent collapse.
The enhanced-label objective is assigned a much smaller weight than the anchoring objective.
Without this weighting, the non-stationarity of intermediate representations can destabilize optimization and hinder convergence.

\section{Experiments}
\subsection{Setup}
We use BEST-RQ \cite{chiu2022bestrq} for unsupervised pre-training and use an
implementation based on \cite{whetten2024bestrq}.
The encoder consists of a VGG front-end and 12 Conformer \cite{gulati2020conformer} blocks.
We use 80-dimensional log-Mel filterbank features with a 10ms frame shift as input.
The features are normalized to zero mean and unit variance, as a lack of normalization
may lead to codebook collapse.
Before inputting the data into the encoder, we randomly mask $\sim$60\% of the
frames in each sequence, where each mask covers a span of 16 frames.
We set the codebook size to 8192 as suggested in \cite{chiu2022bestrq}, and use 256 for codebook distillation since the intermediate representations are closer to phoneme-level and less variable, so a smaller codebook size is already sufficient and more efficient.
The front-end applies a 4× time-wise downsampling.
We set the model size to 512, the number of
attention heads is 8 and the hidden dimension of the feedforward module is
2048.
Following \cite{shaw2018relpe}, we incorporate relative positional encoding into the self-attention.
In practice, we normalize $\mathcal{L}_\text{distill}$ by the sequence length and set the loss scale to 0.5.

We use the full 960h {\LBS} \cite{panayotov2015lbs} audio for pre-training and
train all the models for 30 epochs.
For supervised fine-tuning, we use 10-hour and 1-hour splits from \textit{Libri-light} \cite{kahn2020librilight} and a 100-hour split from {\LBS}.
We fine-tune all models for 30 epochs on the 100-hour split and for 100 epochs on the 10-hour and 1-hour splits.
SpecAugment \cite{wei2020specaug} is applied for data augmentation in
supervised training.
For models trained from scratch, we train with twice the number of epochs for each split.
We use a set of 79 end-of-word augmented phonemes \cite{zhou2021eowphoneme} as prediction targets and train with Connectionist Temporal Classification (CTC) \cite{graves2006ctc}, both for fine-tuning and training from scratch.
All experiments are implemented in RETURNN \cite{zeyer2018returnn}, which we use for both pre-training and supervised training.
During inference, we perform Viterbi decoding with a 4-gram word-level language model using RASR \cite{wiesler2014rasr}.
For incremental PCA implementation, we use PCAonGPU\footnote{https://github.com/dnhkng/PCAonGPU}.
The config files
and code to reproduce the results can be found online\footnote{https://github.com/rwth-i6/returnn-experiments/tree/master/2026-enhance-bestrq}.

\subsection{Overall results}
Table \ref{tab:overall_comparison} shows the effect of applying the proposed methods
incrementally during pre-training on ASR performance.
The results show that fine-tuning from a pre-trained model yields substantial
relative improvement over training from scratch, with larger gains observed when
less supervised data is available.
We run five random seeds for the baseline and choose the best-performing seed.
To control randomness and ensure comparability, we use this fixed seed for all
subsequent experiments.
Moreover, our proposed methods can be combined to provide incremental improvements,
with each method contributing roughly 3–4\% relative improvement.
With all three methods together, we achieve $\sim$12\% relative improvement, i.e.
from 10.1\% to 8.8\% on {\LBS} test-other compared to fine-tuning from the original BEST-RQ pre-trained model.
Moreover, we compare against BiRQ \cite{jiang2025birq}.
Following their setup, we use 0.1 for the enhanced-label objective and 2.4 for
the anchoring objective.
Under these weights, adding enhanced-labels yields a consistent but modest improvement,
likely because the enhanced-label term contributes only weakly to the overall
optimization.

\begin{table}[h!]
\caption{ASR results on the {\LBS} dev/test sets comparing models trained from
scratch or fine-tuned with various pre-trained models using varying amounts of
supervised data. The proposed methods can be applied incrementally in pre-training,
with separate results reported after each addition.}
\label{tab:overall_comparison}
\centering
\renewcommand{\arraystretch}{1}
\setlength{\tabcolsep}{0.25em}
\scalebox{0.85}{\begin{tabular}{|>{\raggedright\arraybackslash}p{5cm}|c|c|c|c|}
\hline
\multirow{2}{*}{Model} & \multicolumn{4}{c|}{WER (\%)} \\ \cline{2-5}
 & dev-cln & test-cln & dev-o & test-oth \\ \hline

\textbf{100 hr labeled data} & & & & \\
\hspace{1em}train from scratch & 4.5 & 5.0 & 14.6 & 14.7 \\
\hspace{1em}fine-tune BEST-RQ baseline & 3.9 & 4.2 & 10.0 & 10.1 \\

\hspace{2em}+ PCA projection & 3.7 & 4.1 & 9.7 & 9.5 \\
\hspace{3em}+ Iterative CB refinement & \textbf{3.5} & 4.0 & 9.1 & 9.2 \\
\hspace{4em}+ CB distillation & 3.6 & \textbf{3.9} & \textbf{8.9} & \textbf{8.8} \\
\hspace{1em}BiRQ \cite{jiang2025birq} & 3.8 & 4.2 & 9.7 & 9.8\\
\hline

\textbf{10 hr labeled data} & & & & \\
\hspace{1em}train from scratch & 17.6 & 17.8 & 38.7 & 39.2\\
\hspace{1em}fine-tune BEST-RQ baseline & 5.3 & 5.7 & 12.7 & 12.6  \\
\hspace{2em}+ PCA projection & 5.3 & 5.7 & 12.6 & 12.6 \\
\hspace{3em}+ Iterative CB refinement & 5.0 & 5.5 & 11.6 & 11.7  \\
\hspace{4em}+ CB distillation & \textbf{4.9} & \textbf{5.1} & \textbf{11.2} & \textbf{11.2} \\
\hspace{1em}BiRQ \cite{jiang2025birq} & 5.3 & 5.6 & 12.2 & 12.4 \\
\hline

\textbf{1 hr labeled data} & & & & \\
\hspace{1em}train from scratch & 90.5 & 90.1 & 93.4 & 93.0\\
\hspace{1em}fine-tune BEST-RQ baseline & 7.7 & 8.1 & 16.6 & 16.9\\
\hspace{2em}+ PCA projection & 7.8 & 8.0 & 16.5 & 16.6 \\
\hspace{3em}+ Iterative CB refinement & 7.2 & 7.5 & 15.0 & 15.1 \\
\hspace{4em}+ CB distillation & \textbf{7.0} & \textbf{7.1} & \textbf{14.7} & \textbf{14.8} \\
\hspace{1em}BiRQ \cite{jiang2025birq} & 7.3 & 7.7 & 16.0 & 16.0\\
\hline
 \end{tabular}}
\vspace{-1.5em}
\end{table}

\subsection{Comparison with using multiple codebooks}
Table \ref{tab:trainin_time_comparison} compares training time per epoch in the
SSL phase and downstream ASR performance when pre-training with multiple codebooks.
The results indicate that using more random codebooks improves downstream ASR
performance, with gains plateauing after four codebooks.
In terms of training cost, adding multiple codebooks introduces significant overhead
(30\% more for 4 codebooks, 52\% more for 6 codebooks).

We find that applying PCA projection and iterative codebook refinement to
a single codebook substantially improves the fine-tuned WER by $\sim$9\% relative, from 10.1\% to 9.2\%,
on {\LBS} test-other, achieving performance comparable to using six random codebooks but incurring minimal
additional cost.
These methods appear to mitigate the effects of poor random initialization,
similar to adding more codebooks that yield more robust pseudo-labels.
When applied to a pre-trained model with two codebooks, the improvement is modest,
and adding more than two codebooks offers little further benefit.
We hypothesize that, for multiple codebooks, despite differing initialization,
the distribution of codebook entries is updated to become closer.
We analyze this further in Sec.~\ref{sec:emd_analysis}.

With PCA projection and a single codebook using iterative codebook refinement,
adding another codebook updated in the same way provides no further gain.
In contrast, adding a codebook updated via codebook distillation yields an
additional 3–4\% relative WER improvement.
This indicates that the extra pseudo-labels encode temporal structure from richer intermediate layer representations, providing more informative supervision.

\begin{table}[h!]
\setlength{\tabcolsep}{1pt} 
\caption{Comparison of training time and WERs using various down projection methods, codebook sizes, codebook update methods for the quantizer. Training time is
reported in hours for 1 epoch using single H100 GPU.}
\label{tab:trainin_time_comparison}
\centering
\scalebox{0.88}{\begin{tabular}{|c|C{3.4cm}|c|c|c|c|c|}
\hline
\multirow{2}{*}{\makecell{Down\\projection}} & \multirow{2}{*}{\makecell{CB update \\ method}} & \multirow{2}{*}{\# CB} & \multirow{2}{*}{\makecell{Train time \\ 1 epoch \texttt{[h]}}} & \multicolumn{2}{c|}{WER} \\
& & & & dev-o & test-o \\ \hline
\multirow{4}{*}{\makecell{random \\ initialized \\ and fixed}} & \multirow{4}{*}{\centering \makecell{random \\ initialized \\ and fixed}} & 1 & 1.31 & 10.0 & 10.1 \\ \cline{3-6}
& & 2 & 1.42 & 9.3 & 9.6 \\ \cline{3-6}
& & 4 & 1.71 & 9.0 & 9.2 \\ \cline{3-6}
& & 6 & 1.99 & 9.1 & 9.2 \\ \hline
\multirow{6}{*}{PCA} & \multirow{4}{*}{\makecell{Iterative CB\\refinement}} & 1 & 1.31 & 9.1 & 9.2 \\ \cline{3-6}
 & & 2 & 1.42 & 9.1 & 9.1 \\ \cline{3-6}
 & & 4 & 1.75 & 9.0 & 9.2 \\ \cline{3-6}
 & & 6 & 2.05 & \textbf{8.9} & 9.1 \\ \cline{2-6}
 & \makecell{
  \underline{Iterative CB refinement} \\
  + \uuline{CB distillation}
} & \makecell{2\\(=\underline{1}+\uuline{1})} & 1.56 & \textbf{8.9} & \textbf{8.8}  \\ \hline
\end{tabular}}
\vspace{-1.5em}
\end{table}

\subsection{Dependence on the initialization}
\label{sec:emd_analysis}
\begin{figure}
  \centering
  \includegraphics[scale=0.23]{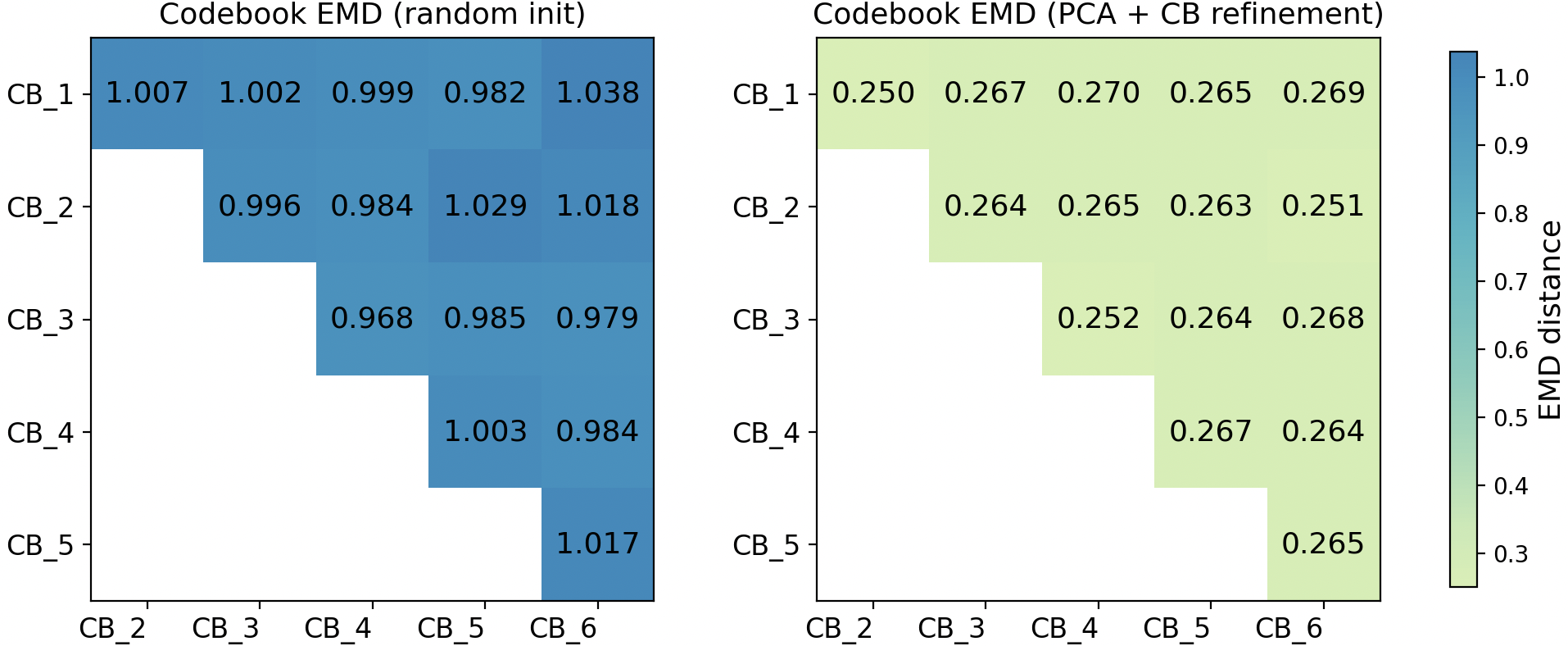}
  \caption{Pairwise EMD distance between randomly initialized codebooks and
  codebooks with PCA and iterative codebook refinement.}
  \label{fig:EMD_distance.png}
  \vspace{-0.5em}
\end{figure}

To assess whether PCA and iterative codebook refinement reduce sensitivity to random initialization, inspired by \cite{lu2023ot}, we measure the Earth Mover’s Distance (EMD) between codebook distributions via optimal transport with Euclidean cost.
EMD is a metric that measures the minimum cost transport distance required to transform one distribution into another.
We use the frequencies of pseudo-label indices in the {\LBS} dev set as the marginal distribution.
At the start of training, the six codebooks are independently initialized at random and therefore differ substantially.
As shown in Figure~\ref{fig:EMD_distance.png}, their pairwise EMDs are large (average 0.99).
After applying PCA and iterative refinement, the average pairwise EMD drops to 0.25, indicating that the learned codebooks converge to a similar configuration despite different initializations.

\subsection{Ablation study on iterative codebook refinement}
Table \ref{tab:ablation_study_centroid2cb} examines the effect of the number of
steps between iterative codebook refinement updates on downstream ASR performance.
Overall, the differences in downstream performance are minimal. However, we observe
that too infrequent updates may not sufficiently refine the pseudo-labels.

\begin{table}[h!]
\setlength{\tabcolsep}{1pt} 
\centering
\caption{Ablation study on the effect of iterative codebook refinement update frequency
(i.e., every \# of training steps) on downstream ASR WER.}
\label{tab:ablation_study_centroid2cb}
\scalebox{0.9}{\begin{tabular}{|c|c|c|c|c|}
\hline
\multirow{2}{*}{\# update steps} & \multicolumn{4}{c|}{WER} \\ \cline{2-5}
 & dev-clean & test-clean & dev-other & test-other \\ \hline
 3600 & 3.5 & 4.0 & 9.1 & 9.2 \\ \hline
 5400 & 3.6 & 4.0 & 9.3 & 9.2 \\ \hline
 7200 & 3.6 & 4.0 & 9.2 & 9.3 \\ \hline
\end{tabular}}
\vspace{-0.5em}
\end{table}

\subsection{Ablation study with codebook distillation}
In Table \ref{tab:ablation_study_distill2cb}, we examine how varying the
distillation layer and feature assignment
type (hard vs. soft weighted centroid) affects fine-tuned WER when using
codebook distillation in pre-training.
The results indicate that using soft or hard assignment types has little impact
on performance. Distilling from intermediate layers, such as \{5,6,7\} or \{6,7,8\},
yields the best results.
For combined layers like \{5,6,7\}, we compute the temporal self-similarity matrix for
each layer and use the averaged matrix for distillation.
As already observed in \cite{hsu2021hubert,pasad2021analysis,baeski2021unsupervised}, intermediate layers of the
pre-trained model provide the most informative features for downstream ASR, as
they capture a balance of phonetic detail and contextual abstraction.
Lower layers focus on low-level acoustic details and lack sufficient contextual
information, while higher layers are tuned to the pre-training objective and may lose
fine-grained phonetic cues.

\begin{table}[h!]
\centering
\caption{Fine-tuned WER (\%) for different distillation layers and feature assignment types using codebook distillation in pre-training.}
\label{tab:ablation_study_distill2cb}
\scalebox{0.9}{\begin{tabular}{|c|c|c|c|}
\hline
\multirow{2}{*}{\makecell{Assignment type}} & \multirow{2}{*}{\makecell{Distill layers}} & \multicolumn{2}{c|}{WER} \\ \cline{3-4}
 &  & dev-other  & test-other \\ \hline


 \multirow{2}{*}{soft} & \{7,8,9\} & 9.1 & 9.0 \\ \cline{2-4}
  & \multirow{2}{*}{\{6,7,8\}} & 8.9 & 8.8 \\ \cline{1-1} \cline{3-4}
 \multirow{6}{*}{hard} & & 8.9 & 8.8 \\ \cline{2-4}
  & \{5,6,7\} & 8.8 & 8.9 \\ \cline{2-4}
  & \{4\} & 9.0 & 9.0 \\ \cline{2-4}
  & \{5\} & 8.8 & 8.9 \\  \cline{2-4}
  & \{6\} & 8.9 & 9.0 \\  \cline{2-4}
  & \{7\} & 8.8 & 8.9 \\ \hline

\end{tabular}}
\vspace{-1em}
\end{table}

\section{Conclusion}
In this work, we introduced three effective approaches to enhance the quantizer
in BEST-RQ and refine the generation of pseudo-labels.
Our experimental results demonstrate that these optimizations achieve a $\sim$12\% (from 10.1\% to 8.8\%)
relative improvement on {\LBS} test-other while adding minimal additional pre-training cost.
These findings highlight that improving the quantizer directly leads to
higher-quality pseudo-labels and stronger learned representations, offering a
simple yet powerful strategy to boost both the efficiency and effectiveness of
self-supervised speech representation learning.

\newpage
\section{Acknowledgments}
This work was partially supported by the project RESCALE within the program \textit{AI Lighthouse Projects for the Environment, Climate, Nature and Resources} funded by the Federal Ministry for the Environment, Nature Conservation, Nuclear Safety and Consumer Protection (BMUV), funding IDs: 67KI32006A and 67KI32006B.

\section{Generative AI Use Disclosure}
Generative AI tools were used for language editing, grammar correction, code assistance, and improving the readability of the manuscript. All technical content, experiments, analyses, and conclusions were produced and verified by the authors.

\bibliographystyle{IEEEtran}
\bibliography{mybib}

\end{document}